# Minimizing Shortfall

Lisa R. Goldberg, Michael Y. Hayes, Ola Mahmoud

## 1 Introduction

Despite the increasing sophistication of Finance in the past 30 years, quantitative tools for building portfolios remain entrenched in the paradigm proposed by Markowitz in 1952; these tools offer investors a trade-off between mean return and variance. However, Markowitz himself was not satisfied with variance, which penalizes gains and losses equally. Instead, he preferred semi-variance, which penalizes only losses.

The endurance of mean-variance optimization may be explained, at least in part, by the difficulty of developing a viable alternative. There is a substantial literature devoted to extending or replacing mean-variance optimization and it revolves around three interrelated themes. The first theme concerns the characterization of a useful measure of risk. Axiomatic approaches to this issue in Artzner et al. (1999), Föllmer and Schied (2002) and Föllmer and Schied (2004) indicate that *convexity* is an essential feature of a risk measure. One reason is that portfolio construction based on a non-convex risk measure can lead to a solution that is locally optimal but not globally optimal. Variance, which measures the squared average dispersion of a portfolio return distribution, is convex. Increasingly, the dominant convex measure of downside risk is expected shortfall,[1] which is the expected loss given that a value at risk threshold is breached. An important feature of expected shortfall is its flexibility: by varying the value at risk threshold, expected shortfall can be made sensitive to different parts of the return distribution. Further, even at relatively low confidence levels, expected shortfall probes the entire tail of loss distribution.

The second theme concerns the development of practical algorithms that incorporate alternative risk measures. Portfolio optimization against variance is a tractable quadratic programming problem. An innovation by Rockafellar and Uryasev (2000) makes portfolio optimization against expected shortfall technically practical by formulating it as a linear programming problem.

The efficacy of any quantitative portfolio construction method depends on the accuracy of the estimated inputs, and the third theme in the portfolio construction literature is estimation accuracy. Variance at horizons up to a month can be forecast with accuracy that is sufficient for the purpose of optimization. In contrast, estimates of expected shortfall are typically inappropriate for optimization; data scarcity leads to unreasonably wide error bars, even at a horizon of one day. However, this issue can be addressed for equities with Factor-Based Extreme Risk (FxR).which is an empirical, fundamental factor-based model that forecasts expected shortfall at horizons up to one month. FxR expected shortfall forecasts reflect persistent characteristics of equities, such as the higher asymmetry and downside risk of Growth stocks compared to Value stocks.

In this paper, we combine the innovations described above in an empirical study of expected shortfall optimization with Factor-Based Extreme Risk. We avoid the issue of forecasting mean return by comparing minimum expected shortfall to minimum variance portfolios.[2] Our study is carried out for the US, UK, and Japanese equity markets and it uses Barra Style Factors (Value, Growth, Momentum, etc.). We show that minimizing expected shortfall generally improves performance over minimizing variance, especially during down-markets, over the period 1985-2010. The outperformance of expected shortfall is due to intuitive tilts towards protective factors like Value, and away from aggressive factors like Growth and Momentum. The outperformance is largest for the expected shortfall at relatively low confidence levels, which measures distributional asymmetry rather than the extreme losses.

---

[1] Conditional value ati risk (CVar) and Expected tail loss (ETL) are synonyms for expected shortfall.

[2] An empirical portfolio construction study that has substantial overlap with our study and incorporates forecasts of mean return is in Bender et al. (2010).

# 2 Background

In this section we review the definition and motivation for expected shortfall, the formulation of variance and expected shortfall optimization problems, and the Factor-Based Extreme Risk model.

## *2.1 Volatility and Expected Shortfall as Risk Measures*

Volatility, or the square root of variance, measures the average dispersion over the entire distribution of portfolio gains and losses. Volatility is the central concept in many standard statistics such as risk contribution, beta, and correlation (Goldberg, et al, 2010). Its usefulness stems from its empirical and mathematical properties. Empirically, volatility is persistent from one period to another; realized volatility in one period is highly correlated with realized volatility in the next. Mathematically, volatility is a convex risk measure, amenable to the tools of convex optimization and analysis (Goldberg and Hayes, 2010). A minimum of a convex risk measure is unique, so once a minimum is found it is guaranteed to be a global minimum.

Although useful, volatility does not describe every aspect of risk. Even as he proposed variance as a risk measure, Markowitz (1952) pointed out that a better risk measure would penalize only losses, and he proposed semi-variance as a desirable alternative. An alternative risk measure that has gained attention in recent years is *expected shortfall*, which is the average (or expected) value of the largest losses. The expected shortfall confidence level specifies the magnitude of these largest losses. For example, the 95% expected shortfall is the average over the 5% largest losses. Given N possible portfolio outcomes, expected shortfall is formally defined as

$$s_p = \frac{1}{N(1-p)} \sum_{i=1}^{N(1-p)} r_{(i)} \qquad (1)$$

where $r_{(i)}$ are the ordered return scenarios and p is the confidence level.[3]

## *2.2 Variance and Expected Shortfall Optimization*

Like variance, expected shortfall is a convex risk measure and can be efficiently minimized. For asset weights w, expected returns $\alpha$, risk aversion $\lambda$, and covariance matrix $\Sigma$, the standard mean-variance problem is:

$$\max_w w'\alpha - \lambda w'\Sigma w \qquad (2)$$

Similarly, the mean-expected shortfall problem is:

---

[3] Here we assume that N(1-p) is an integer. The general formula for continuous outcomes is conceptually similar; see Acerbi and Tasche (2001) for a detailed discussion.

$$\max_w w'\alpha - \lambda s_p(w) \qquad (3)$$

where $s_p(w)$ is the empirical expected shortfall estimator at confidence level p. Rockafellar and Uryasev (2000, 2002), Krokhmal, et al (2002), and Bertsimas. et al (2004) show how to formulate shortfall optimization as a linear program (LP) amenable to standard optimization algorithms. In Appendix C we review the formulation of variance, expected shortfall, and combined variance-expected shortfall optimization in more detail.

## *2.3 The Factor-Based Extreme Risk Model*

Although the promise of alternative risk measures has been recognized since variance was first introduced, the absence of reliable forecasts has been an obstacle to practical applications. Here, we review the two main categories of forecasting models.

Parametric models are widely used to forecast alternative measures of portfolio risk. Given a parametric family of candidate distributions for risk factors, statistical estimation techniques can be applied to determine the best in-sample fit to historical data, and simulations based on the winning distribution are used to generate portfolio risk forecasts. The most familiar parametric model is the Normal (Gaussian) distribution. However, the Normal model reduces to a variance model for portfolios that can be represented as a linear combination of the chosen risk factors. This is because any risk measure[4] is a monotonic function of volatility when returns are normally distributed. Therefore, Normal models of risk do not add additional insight for linear instruments when looking at alternative risk measures. To make use of alternative risk measures, a non-normal risk model is needed, and there are endless possibilities. However, financial data may not be adequate to distinguish between parametric families, and different choices can lead to materially different forecasts.[5]

A non-parametric approach uses historical returns as forecast return scenarios (known in the context of Value at Risk as Historical VaR). While avoiding any specific distributional assumptions, Historical VaR and many other empirical models explicitly assume that historical returns are stationary. In other words, they assume that historical observations are drawn from a common distribution. However, it is widely accepted that this is not the case. For example, volatility and correlations change over time, and this constitutes an argument against historical estimates. Moreover, data constraints often necessitate the use of a relatively short return history (e.g., 1 year) for non-parametric estimation. Certain assets may not have sufficient history (e.g., newly issued equities), or their history may be irrelevant (companies that change from growth to value, small cap to large cap, or from one industry to another).

Factor-Based Extreme Risk (FxR) is a non-parametric model that addresses the issues of volatility and correlation regimes and short or irrelevant data histories. For a portfolio of international equities, FxR uses histories of daily equity risk factor returns to generate a substantial data set that can be used to forecast value at risk, expected shortfall, and other alternative risk measures. Since equity factor return histories in developed markets span several decades, FxR data sets range in size between five thousand and ten thousand observations. An important element of the FxR methodology is covariance scaling: each factor return is updated to the current climate so that the histories used to estimate risk are covariance stationary. Covariance-scaling regularizes the volatility and correlation regimes, so that outliers from different regimes can be evaluated on an equal footing. The FxR model is reviewed in

---

[4] More specifically, any risk measure that is a function of the single-period return distribution.

[5] See, for example, Heyde and Kou (2004).

Appendix A, and Dubikovsky, et al (2010) present broad out-of-sample tests that show how the FxR model is more consistent with market behavior than the Conditional Normal model.

## 3 Uncertainty in Expected Shortfall Forecasts

Before endeavoring to construct portfolios based on alternative measures, a basic question must be answered: can the risk measure be estimated with enough certainty to be useful? Two aspects of forecast uncertainty are precision, measured by estimation error, and predictability, measured by persistence (i.e., do historical returns predict future risk?).

### *3.1 Estimation Error*

Optimized weights are subject to error, even for a perfect risk model, because risk is always estimated with a finite sample. Kondor, et al (2007) explain that estimation error increases with the ratio N/T, where N is the number of assets and T is number of observations. Naturally, estimation error plays a larger role in expected shortfall than in volatility, because a large amount of the input data is only used in aggregate to define the largest losses.

We study the effect of estimation error using data simulated from a standard Normal distribution. Because the true distribution of the simulated variables is known, the true optimal portfolio is also known. We measure estimation error - the deviation between the optimized weights and the true weights - in two ways. First, we compute the average risk of the optimized portfolio divided by the true minimum risk (risk error). Second, we compute the average angle between the optimized weight vector and the true optimal weight vector (weight error). Both of these measure the proximity of the optimized portfolio to the true optimum. The risk error measures the average amount of extra risk that is taken in optimized portfolios due to random fluctuations. Because it is not obvious how much risk error is acceptable, we also compute the weight error, which has a concrete acceptable upper bound. This upper bound is defined by the average weight error of a randomly chosen set of positive weights. If the average weight error is greater than this upper bound, an investor is better off guessing at a random set of weights rather than trying to compute an optimal portfolio. This upper bound is around 35 degrees; we compute an analytical formula for the upper bound in Appendix B. The concept of weight error is illustrated for a two-asset portfolio in Figure 1.

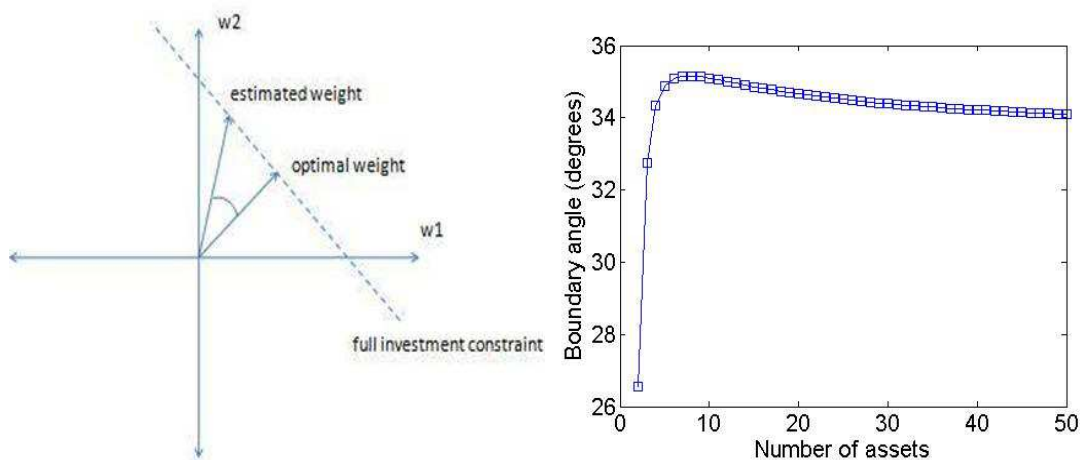

*Figure 1:* Illustration of weight-error angle for a two-asset portfolio (left) and its upper bound (right).

We measure these effects using simulated standard normal random variables applied to 10 assets with sample lengths of 1000, 3000, 5000, and 7000. For each sample length, we simulate returns and minimize expected shortfall at several confidence levels using the full, equal-weighted sample. We compute the ratio of Optimized expected shortfall to true Minimum expected shortfall, we repeat this process 100 times and we average the performance statistics obtained from the 100 repetitions. The results are shown in Figure 2.

We see that for our parameters, in this simplified setting, estimation error is well below the upper bound of 35 degrees that corresponds to random weights. While this does not rule out large estimation error for all possible distributions, our parameters satisfy the baseline normal criterion. We are thus able to control estimation error in this study by considering a low-dimensional risk space (fundamental factors) and a long history of factor returns from 1981.

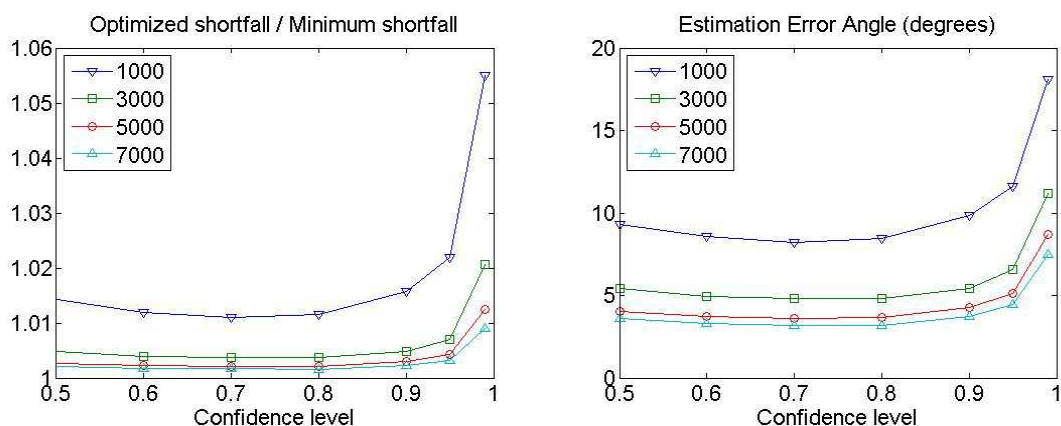

*Figure 2: Measures of estimation error using simulated, standard normal random variables.*

## *3.2 Persistence*

The complementary question to estimation error is persistence. Even if we were able to exactly measure the risk in one period, there would be no use in risk forecasting or portfolio construction if risk did not persist from one period to the next. This is especially relevant for the FxR model, which uses up to 30 years of daily return history to build its forecasts. We postulate that Barra fundamental factors (Size, Growth, Value, etc.) reflect characteristics that are "fundamental" to stock behavior, and are thus persistent across long periods of time. To test this hypothesis, we use the FxR models to forecast expected shortfall from disjoint 15-year samples. Because FxR allows for volatility to evolve over time, we focus on a *non-normality* (NN) statistic that is independent of volatility.[6] The NN statistic is a measure of percent deviation from normality, formally defined as the percent difference between FxR expected shortfall (xShortfall) and a Normal expected shortfall forecast:

$$NN = \frac{xShortfall}{NormalExpectedShortfall} - 1.$$

Positive NN means that the FxR expected shortfall forecast exceeds the normal estimate; negative NN means that the normal estimate exceeds the FxR forecast. Zero NN implies that the FxR and Normal estimates coincide. We compute the NN statistics of Barra USE3 style factors, on FxR estimates of the gain[7] and loss tails, and test the null hypothesis that the NN statistics persist from one period to the next.[8] The results are shown in Figure 3, along with confidence intervals on the difference. If the confidence interval crosses the dashed diagonal, we cannot reject the null hypothesis that the NN statistics persist across periods. Out of 36 factors, we find only a handful of significant outliers: Size at three confidence levels, and Earnings Yield at the 60% confidence level. This result shows that Barra fundamental factors display persistent gain/loss tail features, even over long time periods,. This experiment fails to reject the proposition that suitably modified long histories of fundamental factors are a useful input to portfolio construction. Furthermore, the long history provides a large number of forecast scenarios, allowing us to control estimation error (under the assumption of stationarity) as described in the previous section.

---

[6] More specifically, NN is independent of volatility when Normal expected shortfall is estimated using the same half-life as that used to normalize the FxR returns. Mathematically, NN of a factor is the same as NN of a factor times a positive constant.

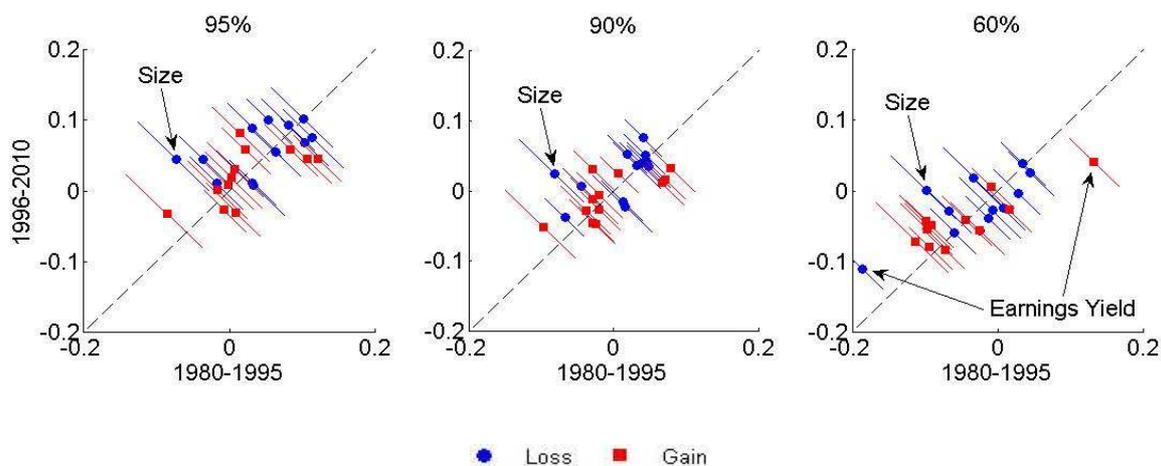

*Figure 3: Persistence of non-normality (NN) of loss/gain tails of style factors in the Barra US factor model (USE3).*

## 4 Optimization Framework

For our empirical study, we consider the active management problem in terms of fundamental factors. We examine the performance of an active strategy that is long minimum expected shortfall and short minimum variance. Alternatively, this performance measures the value added by minimizing expected shortfall instead of minimizing variance.

To simplify the analysis, we do not include expected returns (alpha), and seek only to minimize risk. We further simplify the analysis by considering a small optimization universe consisting of a market index and a small number of Barra Style Factor portfolios (Size, Growth, Value, etc.; see Table 4).[9] We carry out the study in three markets: US, UK, and Japan, during the period 1985-2010.

Each minimum expected shortfall and minimum variance portfolio is constrained as follows: the weight of the index is set to 100% to reflect full investment in the market portfolio. Consequently, the index weight of the active strategy is zero. We constrain each individual style factor exposure to the range [-2, 2]. The sum of style factor exposures is set to zero, so that the active bets are dollar-neutral. This enforces a reasonable level of Style exposure that can be achieved using a moderately sized investment universe.

---

[7] Expected gain is analogous to expected shortfall. It is the average of the largest gains: those exceeding a specified confidence level.

[8] Confidence intervals are computed by bootstrapping the FxR scaled returns in each period.

[9] A factor portfolio return is equivalent to the Barra factor return; some Barra Style factor portfolios are also listed as MSCI indices.

Inputs to the expected shortfall optimizer are daily returns prior to the analysis date. The return history begins in 1981 and the backtesting period starts in 1985, so the expected shortfall optimization is informed by a minimum history of 4 years. These time-series are adjusted using the FxR methodology as explained in Appendix A, representing forecast return scenarios in the expected shortfall objective function. Covariance forecasts are made using an exponentially weighted moving average (EWMA) of trailing factor returns.

We tested a range of parameters, including: the expected shortfall confidence level (60% to 99%); the correlation and volatility half-lives used for FxR covariance rescaling and for the forecast covariance matrix (21, 90, 180 days); and the rebalancing frequency (daily, weekly, monthly, quarterly). With the exception of confidence level, we find little sensitivity to the optimization parameters. Here we focus on a 21-day half-life and a monthly rebalancing frequency.

| US | UK | Japan |
|---|---|---|
| (MSCI USA) | (MSCI UK) | (MSCI Japan) |
| Volatility | Size | Volatility |
| Momentum | Momentum | Size |
| Size | Volatility | Momentum |
| Trading Activity | Trading Activity | Trading Activity |
| Growth | Leverage | Value |
| Earnings Yield | Value | Interest Rate Sensitivity |
| Value | Yield | Growth |
| Earnings Variability | Foreign Sensitivity | Leverage |
| Leverage | Growth | Foreign Sensitivity |
| Currency Sensitivity | | |
| Yield | | |

*Table 4*: Equity style factors and "market factors" (MSCI USA, UK, Japan) used in optimization.

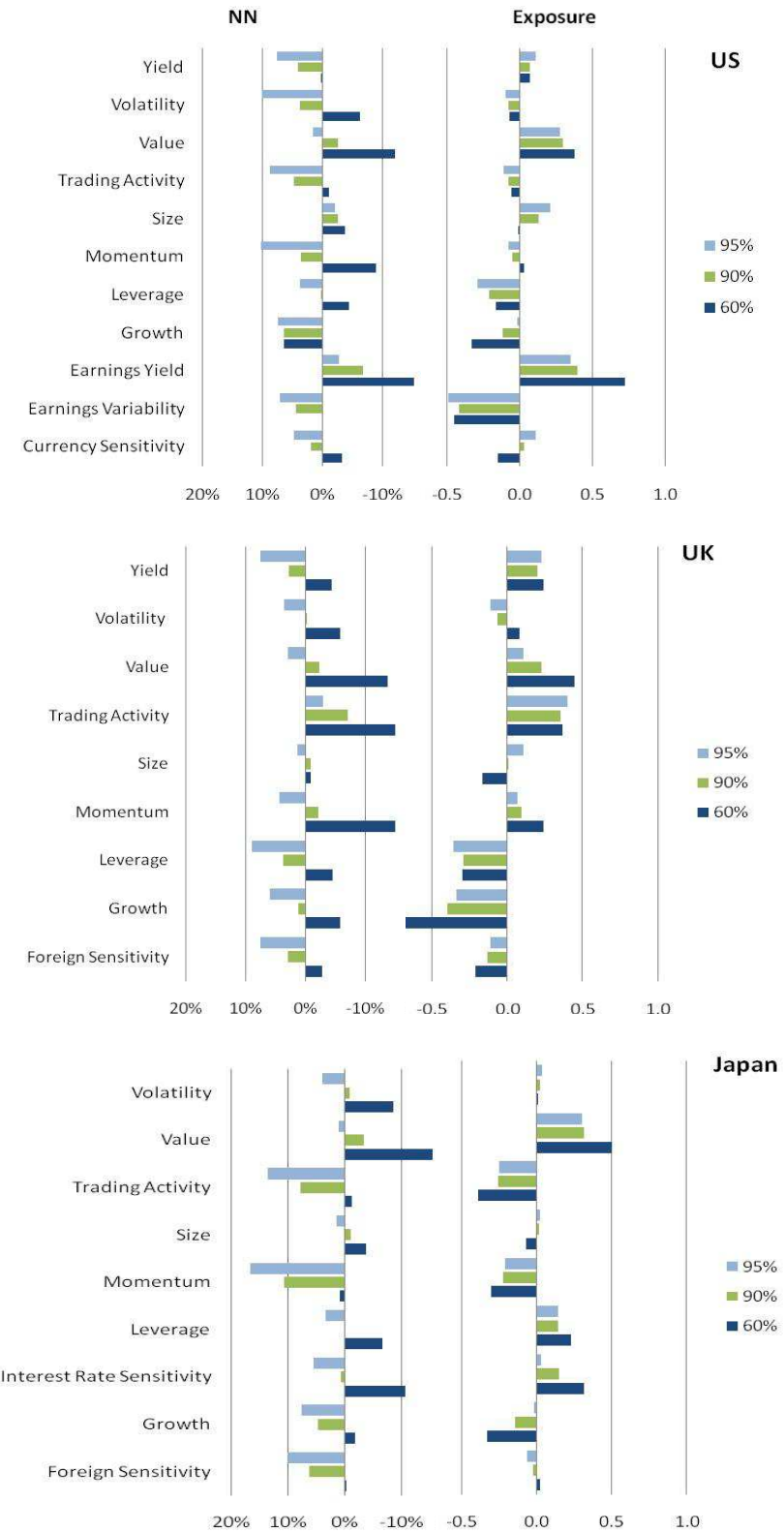

*Figure 5:* Non-normality (NN) of style factors (left) and average monthly excess exposure of minimum expected shortfall over minimum variance (right).

# 5 Optimization Results

## 5.1 Optimal Exposures

Style factor exposures of the active portfolio are shown in Figure 5. The active strategy tilts consistently towards Value in the UK and Japan, and the related Earnings Yield factor in the US. It tilts away from Leverage, Growth, and Earnings Variability in the US, Leverage and Growth in the UK, and Trading Activity, Leverage, and Growth in Japan.

The excess exposures can be partly understood in terms of the NN statistics, also shown in Figures 5 and 6. The factors favored by minimum expected shortfall have smaller (more negative) NN statistics, and those avoided have larger (more positive) NN statistics. The largest visible exception to this trend is 60% Growth in the UK, which is under-weighted in spite of its negative NN. This can be partly explained by the fact that Growth is the fourth riskiest factors in the UK market as measured by NN. Other exceptions include Currency Sensitivity in the US and the related Foreign Sensitivity in the UK and Japan, which can also be explained by their NN relative to the other available factors.

The Volatility, Size, and Trading Activity factors are highly correlated with the market, and are an attractive hedge of market risk in both minimum variance and minimum expected shortfall. Consequently, their exposures consistently approach or coincide with the lower bound of -2. This means that their exposure in the active portfolio is close to zero. Two exceptions are Trading Activity in Japan and the UK. In Japan, Trading Activity is one of the riskiest factors (by NN) and therefore favored by the variance optimizer. In contrast, in the UK, Trading Activity is one of the least risky factors (by NN), so is favored by the expected shortfall optimizer.

FxR captures not only the extremes, but also the overall asymmetry of the distribution. The NN statistics show that at high confidence levels, as is well-appreciated, most factors are riskier than Normal (*fat-tailed*). However, at the 60% level, where expected shortfall examines nearly the entire loss side of the distribution, many factors are in fact *less* risky than Normal (*positively skewed*). For the Barra style factors, the strongest non-Normal risk signal comes from overall asymmetry rather than the extremes. This leads to larger active bets at lower confidence levels.

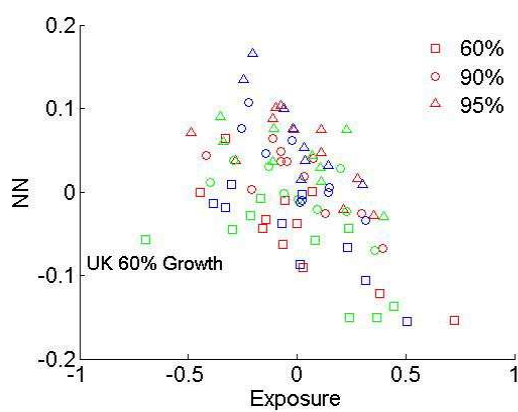

*Figure 6: Relationship of active factor exposure to non-normality (NN); (US – red, UK – green, Japan – blue).*

## 5.2 Optimal Portfolios

### Performance

Figure 7 shows the cumulative returns of the active portfolios. The active strategy shows consistently strong performance over the entire back-testing period for all confidence levels in all three equity markets. In other words, minimum expected shortfall consistently outperforms minimum variance. We also see that the lower the expected shortfall confidence level, the larger the outperformance. It is worth noting that in the 1-2 years leading up to a financial crisis (e.g. 1986, 1998-99), minimum variance outperforms minimum expected shortfall. This is followed by a large improvement in the minimum expected shortfall portfolios during the subsequent turmoil.

When comparing these returns to the market returns, we see that (especially in the US) they are almost mirror images of one another. When a crisis hits, xShortfall optimal excess portfolios remain unaffected and even show gains. This observation suggests that the active portfolios (long minimum expected shortfall, short minimum variance) can be used for downside protection: outperforming the market and limiting losses during turbulent times.

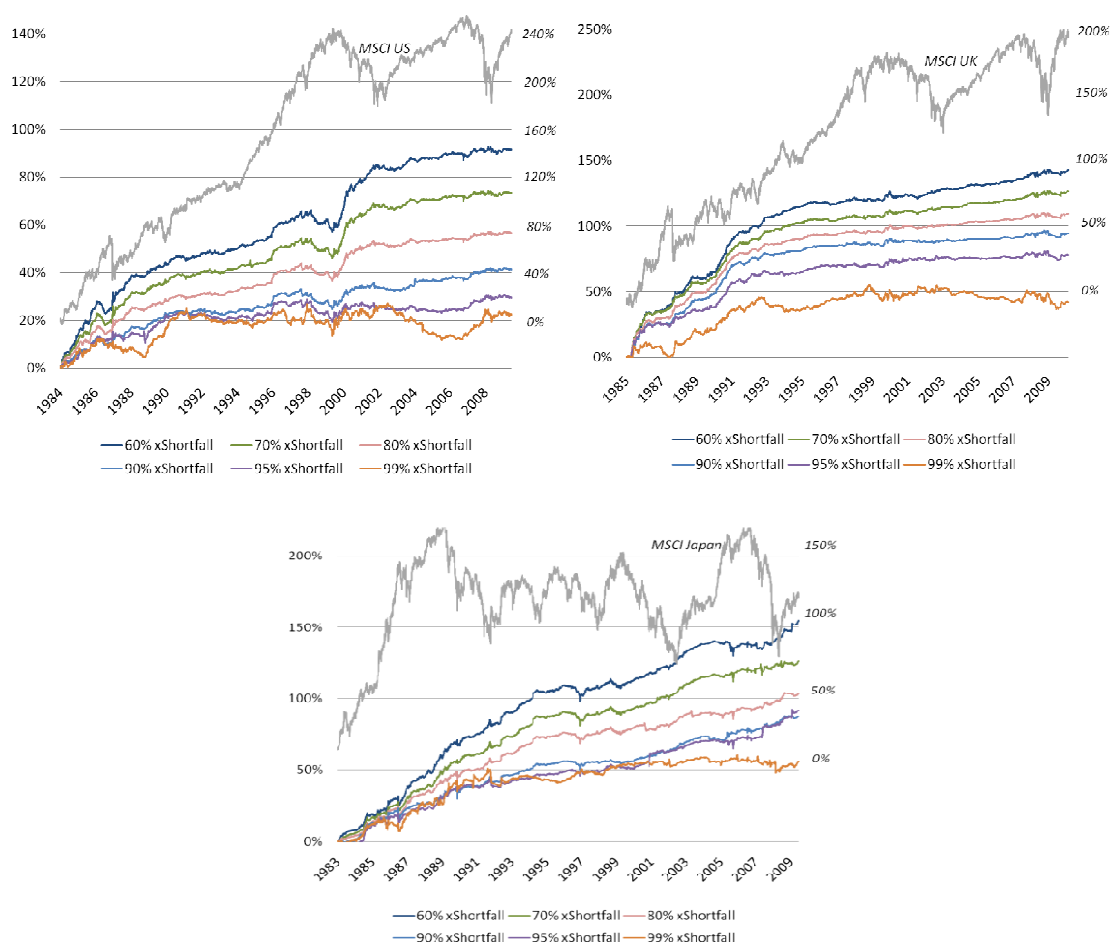

*Figure 7:* Daily cumulative returns of xShortfall-optimal portfolios in excess of variance-optimal portfolio for the US (top left), UK (top right) and Japan (bottom). The corresponding cumulative returns of the market index are shown in grey, with their scale given by the axis on the right.

The outperformance and downside protection that the minimum expected shortfall portfolios offer is also apparent when comparing their absolute cumulative returns to those of the market (see Appendix E). We see that in all three markets the 60% xShortfall optimal portfolio constantly floats above the market. On the other hand, the variance optimal portfolio either exceeds the market (UK) or underperforms, especially in down-markets (US and Japan).

*Return Attribution*

To understand the outperformance of minimum expected shortfall, we perform a return attribution on the excess returns of minimum expected shortfall over minimum variance. Figure 8 shows the cumulative returns for each style factor multiplied by their excess exposures at the 60% expected shortfall confidence level. Ignoring compounding effects, the sum of the cumulative returns of each of these factors equals the excess returns of minimum expected over minimum variance.

Most of the outperformance of minimum expected shortfall is due to tilts towards Earnings Yield (US), Value (UK and Japan), and Trading Activity (UK), and away from Growth (US, UK, Japan), Trading Activity (Japan), Leverage (Japan), and Momentum (Japan). Most of these tilts correspond to conventional wisdom about stock characteristics: Value is protective, and Growth, Leverage, and Momentum are aggressive. A notable exception is Trading Activity, which plays little role in the US, but is favored in the UK and avoided in Japan. This may be partly due to the different definition of Trading Activity in the two models,[10] but more likely reflects qualitative differences between the Japan and UK equity markets. The Japanese equity market has been bearish for most of the test period, while the UK has followed the global business cycle.

In Figure 5 we showed that the excess exposures decrease for higher expected confidence levels. This is because the strongest signal of non-normality comes from examining the entire loss side of the distribution, and not just the tail. Strikingly, the return attribution shows that considering the core of the return distribution will limit losses more effectively than looking deeper into the loss tail.

*Risk Analysis*

Having looked at the performance of minimum expected shortfall portfolios relative to their minimum variance counterparts, a remaining question is whether minimizing expected shortfall reduces extreme risk. To answer this, we compute the realized volatility, realized Sharpe ratio (realized risk/realized volatility), and realized 95% expected shortfall, (average over the largest losses) for down- and up-markets (Table 9). As down-markets, we take the crisis periods of 1987-1988, 2000-2002, and 2007-2008, and up-markets, the remaining years in our back-testing period. When comparing the variance and expected shortfall optimal portfolios in absolute terms, we see that realized volatility is similar for all optimal portfolios, and consistently lower than that of the market. The Sharpe ratio is higher than that of the market, and highest for the 60% expected shortfall optimal portfolio. All optimized portfolios have lower realized expected shortfall than the market does. The active portfolios have negligible realized risk (volatility and expected shortfall) and higher Sharpe ratios, especially during down-markets. We finally see that in the framework of full-investment in the market together with an active hedge using our optimal excess portfolios, realized risk would only decrease by 1%, but realized expected shortfall would be reduced during down-markets. Analysis of the realized risk of the optimal portfolios in the UK and Japan markets shows similar results.

---

[10] Trading Activity in UKE7 is a weighted average of monthly, quarterly, and annual share turnover, while in JPE3 it also includes recent growth in trading volume; however, this descriptor is weighted by only 3%; the remainder of the factor is defined analogously to the UKE7 factor.

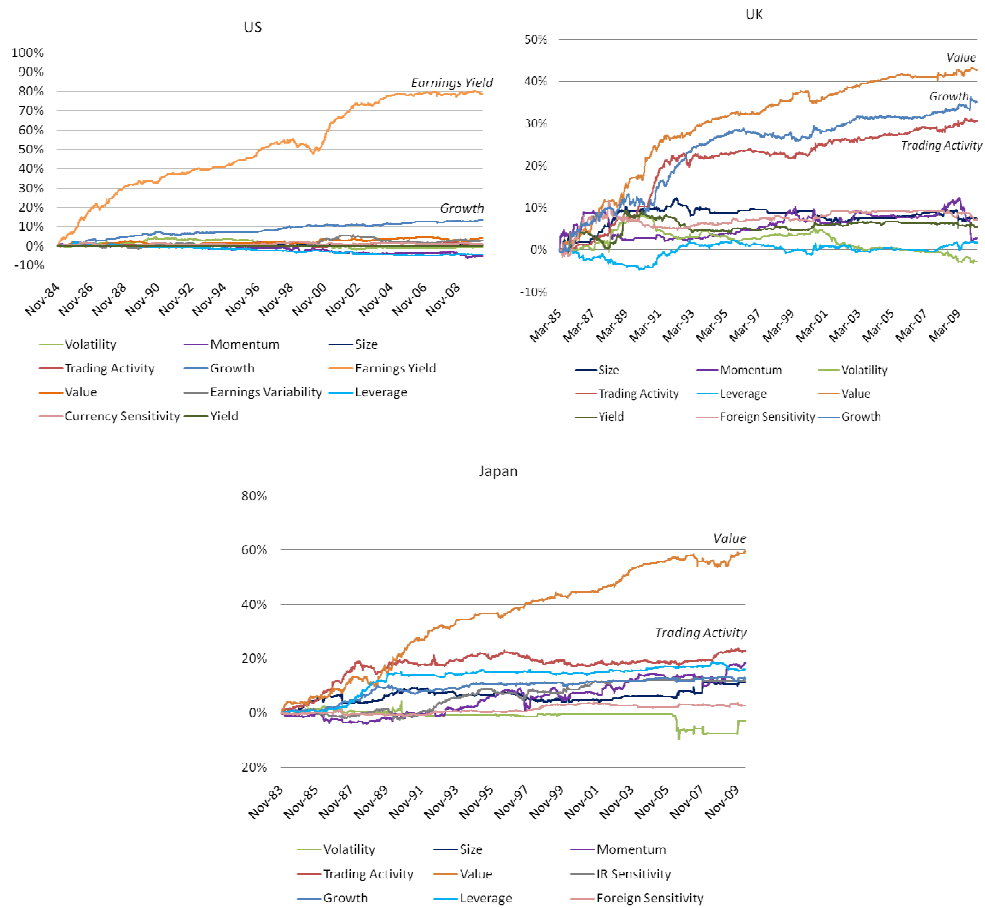

***Figure 8:*** *Return attribution for the monthly rebalanced 60% xShortfall optimal portfolio. For each style factor, its contribution to the overall return is obtained by multiplying its cumulative returns by the excess optimal exposures.*

|  | **Realized Volatility** | | **Realized Sharpe Ratio** | | **Realized 95% Expected Shortfall** | |
| --- | --- | --- | --- | --- | --- | --- |
|  | *Up-markets* | *Down-markets* | *Up-markets* | *Down-markets* | *Up-markets* | *Down-markets* |
| MSCI USA | 15% | 27% | 0.07 | -0.02 | 2.16% | 3.98% |
| Variance-optimal | 11% | 19% | 0.08 | -0.01 | 1.56% | 3.13% |
| 60% xShortfall optimal | 11% | 19% | 0.10 | 0.01 | 1.57% | 3.00% |
| 95% xShortfall optimal | 11% | 20% | 0.08 | 0.00 | 1.59% | 3.15% |
| 60% active portfolio | 1% | 2% | 0.05 | 0.10 | 0.22% | 0.26% |
| 95% active portfolio | 1% | 2% | 0.01 | 0.02 | 0.22% | 0.25% |
| MSCI USA + 60% active | 14% | 26% | 0.09 | -0.02 | 1.94% | 3.78% |
| MSCI USA + 95% active | 14% | 26% | 0.09 | -0.02 | 1.96% | 3.83% |

*Table 9: Comparison of average realized volatility, realized Shape ratio (realized return/realized volatility) and realized 95% expected shortfall.*

The variance- and expected shortfall-optimal portfolio beta with respect to the market is shown in Figure 10, using a two-year rolling window. All optimized portfolios display similar betas over time, which means that the beta of the active portfolio is nearly zero. Beta is on average significantly smaller than one, going as low as 0.2, indicating that the returns of our optimized portfolios do not generally follow market returns.

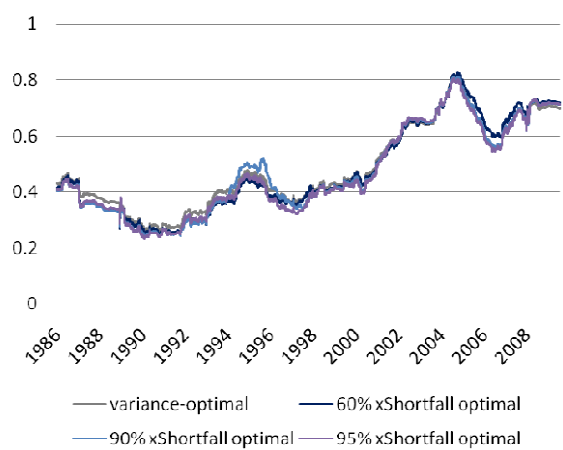

*Figure 10: Evolution of betas for optimal portfolios in the US market. Standard volatility beta with respect to the market is calculated using a two-year rolling window.*

## 6 Conclusion

Our empirical study has shown that expected shortfall optimization, combined with Factor-Based Extreme Risk (FxR), can capture information beyond variance, and that this information can translate to downside protection and superior performance. There are two important empirical observations. The first is that the outperformance of minimum variance by minimum expected shortfall on portfolios of equity style factors is driven primarily by tilts toward Value and Earnings Yield, and tilts away from Growth. Thus, the factor distributions incorporate some of the financial economic effects documented in Fama and French (1992). The second observation is that the outperformance increases as the expected shortfall confidence level decreases. This suggests that it is distributional asymmetry, rather than extreme events, that underlie the results. It also indicates that, in contrast with standard practice, it may be beneficial to place greater emphasis on downside risk measures at relatively low confidence levels.

The novel element of our analysis FxR is a model, which provides a consistent and uniform view of history. Variance has so far been the main risk measure to optimize against due to its simple quadratic definition and its empirical persistence. With the advent of Factor-Based Extreme Risk, downside risk optimization has become a viable alternative to variance optimization.

## Acknowledgements

We thank Carlo Acerbi, Valdislav Dubikovsky, Mark Horvath, Angelo Barbieri, Dan Stefek and Jyh-Huei Lee for their insights and support. We are grateful to two anonymous referees for a careful review that has led to substantial improvements of this article. We also thank the Editorial Staff at *Quantitative Finance* for their time and consideration.

# Appendix A: Factor-Based Extreme Risk Methodology

The non-parametric Factor-Based Extreme Risk (FxR) model forecasts value at risk, expected shortfall, and other alternative risk measures for international equity portfolios. The essential input to the model is a history of daily returns to equity and currency risk factors. For equities, the risk factors are the Barra industry, country and style factors. For currencies, the risk factors are changes in log exchange rates against the US dollar. In developed markets, these return series have histories of up to four decades.

To generate covariance-stationary series, the vector of factor returns on each date is "whitened" by pre-multiplication with the contemporaneous covariance matrix raised to the minus-one-half power. Subsequently, each whitened return is updated to the current covariance regime by multiplication with the current covariance matrix raised to the one-half power.

More precisely, if $\Sigma$ is the covariance matrix of a vector $f$ of factor returns, then $f$ can be expressed as

$$f = \Sigma^{1/2} g \qquad (4)$$

where the ``whitened " vector of factor returns $g$ are uncorrelated and have unit variance. The fundamental assumption of FxR is that the whitened factor returns are identically distributed over time. The uncorrelated factor returns can be recovered from the actual returns with the transformation

$$g = \Sigma^{-1/2} f \qquad (5)$$

In practice, the covariance matrix $\Sigma$ is not known In FxR, we estimate $\Sigma$ using an exponential weighted moving average (EWMA) of trailing factor returns. The halflife is a model variable, and we use a halflife of 21 days in our study.

Forecast scenarios $\tilde{f}_T$ on analysis date T are given by

$$\tilde{f}_T = \Sigma_T^{1/2} \Sigma_t^{-1/2} f_t \qquad (6)$$

where t is a historical date preceding T. These *covariance-normalized factor returns* are the inputs to expected shortfall optimization. Further details are given in Dubikovsky, et al (2010).

## Appendix B: Estimation Error and Derivation of the Weight-Error Angle

We start by mathematically encapsulating the range of all feasible weight vectors **w** that satisfies the given optimization constraints. Consider the two-asset portfolio setting of Figure 1, where the set of all possible weights lies on the intersection of the dashed line with the first quadrant. We define the *feasibility range* in this two-dimensional setting to be the length of this line, which is √2. In three dimensions, feasible weights lie on the area of an equilateral triangle with edge length equal to √2; in four dimensions, feasibility is the volume of a tetrahedron with edge length equal to √2. In general, given n+1 assets and their weights, the feasibility range is the n-dimensional volume of the n-dimensional simplex with equal-sided edges equal to √2. This volume is equal to

$$V_{feasible} = \frac{\sqrt{n+1}}{n!} \qquad (7)$$

Since the assumption is that optimal weights are given by equal weights, the optimal weight vector $\mathbf{w}_{op}$ for n+1 assets has coordinates $\mathbf{w}_{op}$ = [1/(n+1), …, 1/(n+1)], and its length is its norm |**w**|.

Finding the boundary angle β around the optimal weight vector (beyond which one would be better off guessing a random set of weights rather than computing the optimal weights) reduces to finding the n-dimensional object with sub-volume $V_{boundary}$ that has half the volume of the total feasibility range (i.e., $V_{boundary}$ = ½ $V_{feasible}$). Since we seek an angle rotating around the optimal vector $\mathbf{w}_{op}$, the object with volume $V_{boundary}$ must be an n-dimensional hypersphere centered at the coordinates of $\mathbf{w}_{op}$. Its radius is unknown, but can be backed out from its volume:

For n even, $r = \left[\dfrac{\sqrt{n+1}}{2(2\pi)^{n/2}(n!!)}\right]^{1/n}$ and for n odd, $r = \left[\dfrac{\sqrt{n+1}}{4(2\pi)^{n-1/2}(n!!)}\right]^{1/n}$,

where n!!! (respectively n!!) is the product of all odd (respectively even) factors of n. With some higher-dimensional imagination, one may now be able to see that the boundary angle β is adjacent to the optimal vector $\mathbf{w}_{op}$ and has an opposite edge given by the radius r of the n-dimensional hypersphere (with the right angle being between the vector and the radius). Therefore, the boundary angle β is given by

$$\beta = \operatorname{atan}(r/|w_{op}|) \qquad (8)$$

and can be generated for any number of assets n+1 (dimension n) since it is dependent only on n.

## Appendix C: Variance and Expected Shortfall Optimization

In this appendix we review the standard formulation of mean-variance optimization as a quadratic program (QP) and mean-expected shortfall optimization as a linear program (LP). Both QPs and LPs can be solved using standard optimization algorithms.

**Variance Optimization**. Given a vector of weights w, covariance matrix $\Sigma$, vector of expected returns $\alpha$, and risk aversion parameter $\lambda$, the mean-variance optimization problem is:

$$\max_w w'\alpha - \lambda w'\Sigma w \qquad (9)$$

subject to any set of linear equality or inequality constraints (long-only, full investment, etc.).

**Expected Shortfall Optimization**. Given vectors $r_1,\ldots,r_T$ of forecast return scenarios, weight vector w, empirical expected shortfall estimator $s_p(w) = -\frac{1}{K}\sum_{i=1}^{K} w'r_{(i)}$ for confidence level p, $K = \lfloor T(1-p) \rfloor$, and risk aversion parameter $\Lambda$, we seek weights w minimizing $s_p(w)$:

$$\max_w w'\alpha - \Lambda s_p(w) \qquad (10)$$

The expected shortfall estimator $s_p(w)$ is an average over order statistics, and it is not obvious how to solve the expected shortfall optimization. However, Rockafellar and Uryasev (2000) and others have shown how to formulate this as an equivalent linear program with T+1 additional variables and 2T additional constraints:

$$\max_{w,z,t} w'\alpha + \Lambda\left(t - \frac{1}{K}\sum_{i=1}^{T} z_i\right) \qquad (11)$$
$$\text{s.t.} \quad z \geq 0, \quad z_i > t - w'r_i$$

subject to any set of linear equality or inequality constraints (long-only, full investment, etc.). In the next section (Appendix D), we sketch how to convert the optimization problem (8) into its LP equivalent formulation.

**Variance / Expected Shortfall Optimization**. Clearly the variance and expected shortfall terms can be combined into a single objective function to give another standard quadratic program:

$$\max_{w,z,t} w'\alpha + \Lambda\left(t - \frac{1}{K}\sum_{i=1}^{T} z_i\right) - \lambda w'\Sigma w \qquad (12)$$
$$\text{s.t.} \quad z \geq 0, \quad z_i > t - w'r_i$$

subject to any set of linear equality or inequality constraints (long-only, full investment, etc.).

# Appendix D: Linearization of Expected Shortfall Optimization

In this appendix, we begin by considering this expected shortfall minimization problem:

$$\min_w \ s_p(w) \qquad (13)$$

Let $\mathbf{r}_1, \ldots, \mathbf{r}_T$ be T vectors of forecast return scenarios for N assets, and $\mathbf{w}$ be the weight vector. The portfolio return at time t is given by $p_t = \mathbf{w}'\mathbf{r}_t$. If the sorted portfolio returns are written in increasing order as $(\mathbf{w}'\mathbf{r})_{(1)} \leq (\mathbf{w}'\mathbf{r})_{(2)} \leq \cdots \leq (\mathbf{w}'\mathbf{r})_{(T)}$, the empirical expected shortfall estimator is

$$s_p = -\frac{1}{K} \sum_{i=1}^{K} \mathbf{w}'\mathbf{r}_{(i)} \qquad (14)$$

where $K = \lfloor T(1-p) \rfloor$. The linearization of this optimization problem is based on two crucial observations that convert the order statistic into a linear sum that is subject to linear constraints.

**Observation I.** The sum of the t smallest portfolio returns is always smaller than or equal to the sum of any other combination of t returns. Formally, for any $t < T$, we have

$$\sum_{i=1}^{t} \mathbf{w}'\mathbf{r}_{(i)} \leq \sum_{i \in S} \mathbf{w}'\mathbf{r}_i \qquad (15)$$

where we are indexing on the right hand side over all possible sets S that contain t portfolio returns (i.e., $|S|=t$, for $t=1,\ldots, T$). The expected shortfall optimization problem (13) can therefore be rewritten as

$$\begin{aligned} \max_{w,p} \ & -\frac{1}{K}\sum_{i=1}^{K} p_{(i)} \\ \text{s.t.} \ & \sum_{i=1}^{t} p_{(i)} \leq \sum_{i \in S} \mathbf{w}'\mathbf{r}_i \end{aligned} \qquad (16)$$

**Observation II.** The sum of the K smallest portfolio returns $\sum_{i=1}^{K} p_{(i)}$ as it appears in the objective function of (16) is in fact the value of the linear optimization problem (17)

$$\begin{aligned} \min_x \ & \sum_{i=1}^{T} x_i p_i \\ \text{s.t.} \ & \sum_{i=1}^{T} x_i = K, \quad 0 \leq x_i \leq 1 \end{aligned} \qquad (17)$$

This can be proven formally by induction on the number K. However, to help understand why this is true, notice that for K=1 we get the minimum value of combination of portfolio returns if we assign all the weight to the smallest return $p_{(1)}$, since any other fraction of weight assigned to a larger return will yield a larger total value. Similarly, for K=2 we get the minimum combination of portfolio returns if we assign all weights to the smallest returns. Since each $x_i$ cannot be larger than 1, we pick the two smallest returns. By strong duality, optimization problem (17) is equivalent to

$$\max_{t,z} \quad Kt + \sum_{i=1}^{T} z_i$$

$$\text{s.t.} \quad t + z_i \leq p_i, \quad z_i \leq 0, \quad i = 1,\ldots,T \tag{18}$$

We use this observation to rewrite optimization (16) as

$$\min_{w} \quad -\frac{1}{K} \max_{t,z} \quad Kt + \sum_{i=1}^{T} z_i$$

$$\text{s.t.} \quad t + z_i \leq w'r_i, \quad z_i \leq 0, \quad i = 1,\ldots,T \tag{19}$$

Using the fact that max(f) = -min(-f), we finally convert optimization problem (18) into the following linear optimization, which is equivalent to (13):

$$\min_{w,t,z} \quad -t - \frac{1}{K} \sum_{i=1}^{T} z_i$$

$$\text{s.t.} \quad t + z_i \leq w'r_i, \quad z_i \leq 0, \quad i = 1,\ldots,T \tag{20}$$

# Appendix E: Absolute Cumulative Returns

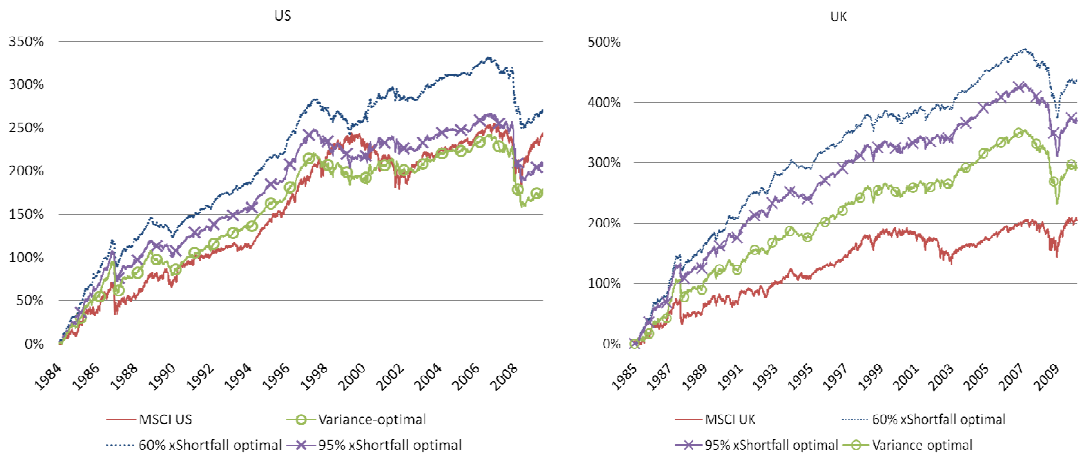

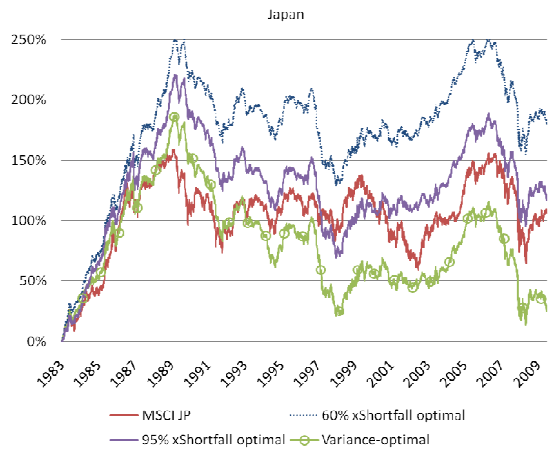